\documentclass[]{mn2e}
\usepackage{graphicx}

\newif\ifAMStwofonts

\def\gs{\mathrel{\raise0.35ex\hbox{$\scriptstyle >$}\kern-0.6em 
\lower0.40ex\hbox{$\scriptstyle \sim$}}}
\def\ls{\mathrel{\raise0.35ex\hbox{$\scriptstyle <$}\kern-0.6em 
\lower0.40ex\hbox{$\scriptstyle \sim$}}}

\newcommand{\be}{\begin{equation}}
\newcommand{\ee}{\end{equation}}
\newcommand{\bea}{\begin{eqnarray}}
\newcommand{\eea}{\end{eqnarray}}

\ifCUPmtlplainloaded \else
  \ifAMStwofonts \else 
    \def\upi{\pi}
                 \def\umu{\mu}
    \def\upartial{\partial}
  \fi
\fi

\title{Substructure and the Cusp and Fold Relations}
\author[Aazami \& Natarajan]{Amir Babak Aazami$^1$, Priyamvada Natarajan$^{1,2}$\\
$^1$Department of Physics, Yale University, P.O. Box 208101, 
       New Haven, CT 06520-8101, USA \\
$^2$Department of Astronomy, Yale University, P.O. Box 208101,
New Haven, CT 06520-8101, USA\\}

\begin{document}

\maketitle

\label{firstpage}

\begin{abstract}
Gravitational lensing of a background source by a foreground galaxy
lens occasionally produces four images of the source. The cusp and the
fold relations impose conditions on the ratios of magnifications of
these four-image lenses. In this theoretical investigation, we explore
the sensitivity of these relations to the presence of substructure in
the lens. Starting with a smooth lens potential, we add varying
amounts of substructure, while keeping the source position fixed, and
find that the fold relation is a more robust indicator of substructure
than the cusp relation for the images. This robustness is independent
of the detailed spatial distribution of the substructure, as well as
of the ellipticity of the lensing potential and the presence of
external shear.
\end{abstract}


\section{Introduction}

Gravitational lensing describes the deflection of light rays from a
background source caused due to the presence of a foreground
concentration of mass (Refsdal 1964; Blandford \& Narayan 1986;
Schneider et al. 1992). The source can be a star, a galaxy, or a
quasar, while the matter concentration, the lens, can also be a star,
an individual galaxy, or a cluster of galaxies.  The geometry of
space-time, defined by a particular cosmological model, is also an
essential ingredient in gravitational lensing, as the strength of the
deflection depends on the relative positions of the source, the lens,
and the observer.  If the lens is a very massive object, such as a
galaxy or a cluster of galaxies, and if it is appropriately aligned
with the background source, then it can produce multiple images of the
source.  This defines the {\lq strong lensing\rq}, regime. In this
work, we study the strong lensing of a background source by a galaxy
lens. The salient feature of gravitational lensing is that this
mapping from the source plane to the image plane preserves surface
brightness.  However, because the monochromatic flux of the source is
usually unobservable, we can only measure the magnification ratios, or
{\it flux ratios}, of the lensed images. Based on these flux ratios,
we can then construct a model of the gravitational lens that will most
accurately reproduce observed values.

Unfortunately, this modeling has turned out to be very difficult in
practice, despite stringent constraints. The primary constraint on
lens modeling comes from the magnification theorem, which states that
for a given source position, the magnification of all images must sum
to at least unity. One of the most important issues in gravitational
lensing is the (growing) number of lenses whose flux ratios {\it
violate} this magnification theorem.  Some of the lenses with
so-called {\lq anomalous flux ratios\rq} are B1422+231, PG1115+080,
B0712+472, B2045+265, and SDSS 0924+0219 (Patnaik et al. 1999, Chiba
et al. 2005, Jackson et al. 1998, Fassnacht et al. 1999, Inada et
al. 2003).  These lenses, among others, exhibit flux ratios that
cannot be fit with smooth lens models. It is reasonable to assume that
these anomalous flux ratios may be due to limitations in our lens
models.  For example, our lens models may have inaccurate or too few
parameters. However, as Mao {\&} Schneider (1998) first demonstrated,
for the case of B1422+231, in which the discrepancy between observed
and model-predicted flux ratios cannot be ascribed to an incorrect
choice of parameterization, but rather to substructure in the lens.

This suggestion is compelling, because it is in accord with
predictions of the granularity of the mass distribution inferred from
high resolution cosmological simulations in a Cold Dark Matter (CDM)
dominated Universe.  These simulations have found copious amounts of
substructure on all scales in the Universe, including galaxy scales
(Diemand et al. 2005; Mathis et al. 2002; Moore et al. 1999). Metcalf
{\&} Madau (2001) quantified the effect of CDM substructure on flux
ratios and found that $10^{4} - 10^{8} M_{\odot}$ substructures near
the Einstein radius can cause the flux ratios to deviate significantly
from their model-predicted values.  Soon afterwards, Dalal {\&}
Kochanek (2002) introduced a method by which to measure the abundance
of substructure using lensing data, and concluded that substructure
comprised $\approx 2\%$ of the mass interior to the Einstein radius of
typical lens galaxies.

Assuming, therefore, that substructures exist and that their effect is
important, we would like to develop a robust diagnostic that
quantifies their presence. In the theoretical investigation presented
in this paper, we keep the source position fixed and study the effect
of substructure on the image configurations.  Aided by previous work
from Keeton et al. (2003, 2005), we have two diagnostics at our
disposal, namely, examining the {\it cusp} and the {\it fold}
relations ($R_{\rm{cusp}}$ and $R_{\rm{fold}}$ hereafter),
which we do in this paper.  We compare the robustness of these two
relations using the publicly available software Gravlens by Keeton
(2001).  We simulate a simple lens model with added substructure, with
the aim of seeing how $R_{\rm{cusp}}$ and $R_{\rm{fold}}$
gauge the presence of substructure as a function of its mass and
position in the lens.

This paper is organized as follows.  In $\S 2$ we give a brief review
of the magnification relations for folds and cusps, and describe the
general properties of $R_{\rm{cusp}}$ and $R_{\rm{fold}}$.
In $\S 3$ we use $R_{\rm{cusp}}$ and $R_{\rm{fold}}$ on a
singular isothermal ellipsoid (SIE).\footnote{In this paper, as in
Gravlens, the ellipticity is defined by $e = 1- q$, where $q$ is the
axis ratio (the ratio of the minor axis to the major axis).}  These
two sections also provide an overview of previous work by Keeton et
al.  In $\S 4$ we add substructure to our lens and investigate the
values of $R_{\rm{cusp}}$ and $R_{\rm{fold}}$ as the masses
and positions of the substructure are varied. We also examine these
relations when the ellipticity of the lens is varied and when external
shear is added to the lens. We present our results and discuss its
implications in $\S 5$.

\section{The cusp and the fold relations}

We begin with a brief review of the necessary lensing terminology.
The lens equation for a gravitational lens system relates the impact
parameter of the source's light ray on the lens plane $L$ to the
source's position on the source plane $S$, by taking into account the
deflection of the ray by the lens mass.  It can be written in
dimensionless form as ${\vec{y}} = {\vec{x}} -
{\vec{\alpha}}({\vec{x}})$, where $\vec{y}$ is the position of the
source, $\vec{x}$ is the source position on the lens plane, and
${\vec{\alpha}}({\vec{x}})$ is the {\it bending angle vector}, which
accounts for the deflection of the light ray.  The lens equation can
also be viewed {\lq in reverse\rq}, as a map $\vec{\eta} : L \to S$
tracing the light ray backwards from the lens to the source plane.
That is, we can view $\vec{\eta}$ as the assignment $\vec{x} \mapsto
\vec{\eta}({\vec{x}}) = {\vec{x}} - {\vec{\alpha}}({\vec{x}})$.  The
inverse of the determinant of the Jacobian of this map,
$({\rm{det}}[{\rm{Jac}}{\vec{\eta}}]({\vec{x}}))^{-1}$, for a lensed
image at position $\vec{x}$ on the lens plane, gives the magnification
of that image, and is conventionally referred to as the {\it
amplification matrix $A({\vec{x}})$}.  The magnification is a signed
quantity: when it is negative, the image is called a {\it saddle;}
when it is positive, it is called a {\it minimum} or a {\it
maximum}.\footnote{The terms {\lq saddle\rq}, {\lq minimum\rq}, and
{\lq maximum\rq} are standard in Morse theory.  The number of minus
signs appearing across the diagonal of the Hessian matrix of a Morse
function gives the number of {\lq downhill\rq} directions of that
function.  In particular, if there are no minus signs, then there are
no downhill directions, and hence the function has a {\it minimum} at
that point.  See Petters et al. (2001) for a comprehensive treatment
of the use of Morse theory in gravitational lensing.}  From the
definition of magnification that those positions $\vec{x}$ for which
${\rm{det}}[{\rm{Jac}}{\vec{\eta}}]({\vec{x}}) = 0$ formally have an
infinite magnification.  The collection of all such points on the lens
plane defines the {\it critical curve}.  The corresponding collection
of points on the source plane defines the {\it caustic curve}.
Focusing on the source plane, the smooth portions of the caustic curve
are called its {\it folds}, while the points where two abutting folds
coincide are called its {\it cusps}.  Typical examples of critical
curves, caustic curves, folds, and cusps are shown in Fig.~1.  For a
detailed treatment of these concepts, see Blandford {\&} Narayan
(1986), Schneider et al. (1992), and Petters et al. (2001).

For folds and cusps, certain local relations between the
magnifications of the multiple images are satisfied (Blandford {\&}
Narayan 1986; Schneider {\&} Weiss 1992; Schneider et al. 1992;
Petters et al. 2001; Gaudi {\&} Petters 2002a,b; Keeton et al. 2003,
2005).  For example, when the source lies asymptotically close to a
cusp caustic (see Fig.~1), the three closely-spaced images (the
so-called cusp triplet) should satisfy $|\mu_{A}| - |\mu_{B}| +
|\mu_{C}| \approx 0$, where $\mu_{i}$ is the signed magnification of
image $i$.  For four-image lenses, Keeton et al. (2003) used this
relation for the cusp triplet to define \bea && R_{\rm{cusp}}
\equiv {|\mu_{A}| - |\mu_{B}| + |\mu_{C}| \over |\mu_{A}| + |\mu_{B}|
+ |\mu_{C}|} = {F_{A} - F_{B} + F_{C} \over F_{A} + F_{B} + F_{C}},
\eea where $F_{i} = F_{\rm{src}}|\mu_{i}|$ is the flux of image
$i$ if the source has flux $F_{\rm{src}}$ (we have essentially
divided out by $F_{\rm{src}}$ since it is unobservable, and are
left with a dimensionless quantity).  $\mu_{B}$ is the magnification
of the middle image, and there is no need to specify whether it is a
minimum or a saddle (for an SIE, it is a saddle if the source lies on
the long axis of the caustic, and a minimum if the source lies on the
short axis).  The ideal cusp relation $R_{\rm{cusp}} \to 0$ is
satisfied only when the source lies asymptotically close to the cusp
caustic.  To move beyond the asymptotic regime, Keeton et al. (2003)
expanded the lens mapping in a Taylor series about the cusp to get
\bea && R_{\rm{cusp}} = 0 + A_{\rm{cusp}}d^{2} + \cdots,
\eea where $d$ is the maximum separation between the three images and
the coefficient $A_{\rm{cusp}}$ is a function that depends on
properties of the lens potential at the cusp point.  In fact, Keeton
et al. (2003) found that the properties of the lens that matter for
$A_{\rm{cusp}}$ are the ellipticity, higher-order multipole
modes, and external shear, whereas the radial mass profile is
unimportant.  Looking at eqn.(2), we see that as the source moves a
small but finite distance from the cusp, $R_{\rm{cusp}}$ picks up
a term to second order in $d$.  Keeton et al. (2003) derived reliable
upper bounds on $R_{\rm{cusp}}$, and concluded that these bounds
would be violated {\it only if the lens potential has significant
structure on scales smaller than the distance between the images}.


\begin{figure}
\begin{center}
\includegraphics[width=8cm]{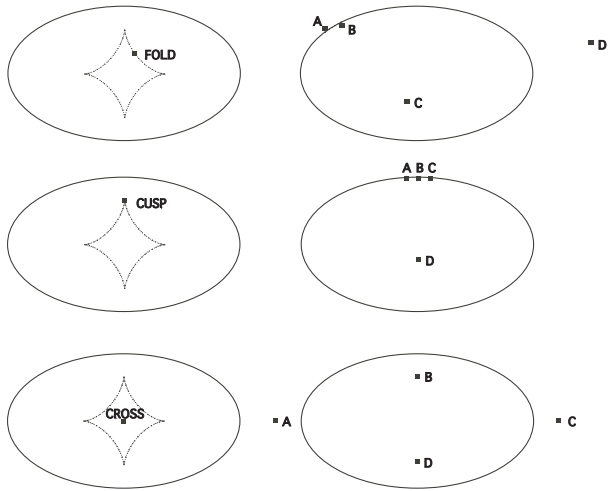}
\caption{The three types of image configurations: folds (top right),
cusps (middle right), and crosses (bottom right), with the
corresponding source position and caustic curve to the left.  The lens
potential used here is an SIE with ellipticity $e = 0.5$.}
\end{center}
\end{figure}

A similar magnification relation holds when the source lies near a
fold caustic (see Fig.~1). In this case, two of the four images lie
closely-spaced together, straddling the critical curve (the so-called
fold image pair).  One of these images is a minimum and the other a
saddle.  When the source lies asymptotically close to the fold
caustic, the fold image pair should satisfy $|\mu_{\rm{min}}| -
|\mu_{\rm{sad}}| \approx 0$.  For four-image lenses, Keeton et
al. (2005) used this relation to define
\begin{eqnarray}
&& R_{\rm{fold}} \equiv {|\mu_{\rm{min}}| - |\mu_{\rm{sad}}| \over |\mu_{\rm{min}}| + |\mu_{\rm{sad}}|} = {F_{\rm{min}} - F_{\rm{sad}} \over F_{\rm{min}} + F_{\rm{sad}}}\cdot
\end{eqnarray}
Like its predecessor, the ideal fold relation $R_{\rm{fold}} \to
0$ is satisfied only when the source lies asymptotically close to the
fold caustic.  To move beyond the asymptotic regime, these authors
expanded the lens mapping in a Taylor series about the fold point to
get 
\bea && R_{\rm{fold}} = 0 + A_{\rm{fold}}d_{1} + \cdots,
\eea 
where $d_{1}$ is now the distance between the two images in the
fold image pair and the coefficient $A_{\rm{fold}}$ is a function
that depends once again on those properties of the lens potential
listed above.  As we approach a cusp in the asymptotic regime,
$|A_{\rm{fold}}| \to \infty$, where the sign depends on whether
the cusp is on the long or short axis of the caustic.  For a cusp on
the long axis, $A_{\rm{fold}} \to -\infty$ because the middle
image is a saddle, whereas for a cusp on the short axis,
$A_{\rm{fold}} \to +\infty$ because the middle image is a minimum
(remember that we are now looking at a cusp triplet, but using
$R_{\rm{fold}}$ instead of $R_{\rm{cusp}}$, and that
$R_{\rm{fold}}$ is to be evaluated on minimum/saddle {\it pairs}
of images).  This means that $R_{\rm{fold}}$ breaks down
asymptotically close to a cusp, which is not too surprising because it
is designed to be evaluated only on fold points.  We mention this fact
because in our calculations below we do indeed evaluate
$R_{\rm{fold}}$ for cusp points, but we are safe in doing so
because our source sits a small but finite distance from the cusp, and
is not asymptotically close.

The more interesting feature of $R_{\rm{fold}}$, as Keeton et
al. (2005) discovered, is that the validity of the ideal fold relation
depends not just on how close the source is to the fold caustic, but
also where precisely the source is {\it along} the caustic.  This is
reflected in the coefficient $A_{\rm{fold}}$, which takes on all
values, both positive and negative, as one moves along the caustic
(see Fig.~2).  For this reason the authors introduced another
variable, the distance $d_{2}$ to the next nearest image (note that,
as we are dealing with minimum/saddle image pairs only, neither
$d_{1}$ nor $d_{2}$ will ever denote the distance between two saddles
or two minima).  They did so because both the source and the fold
caustic are of course unobservable, and therefore the value of
$A_{\rm{fold}}$, too, is unknown.  Fortunately, the source's
position is encoded in the image configuration: not in the separation
$d_{1}$ between the fold image pair, but rather in the distance
$d_{2}$ to the next nearest image.  (To give an example: a source near
a fold but not near a cusp will have $d_{1} \ll d_{2} \sim
R_{\rm{E}}$, whereas a source near a cusp will have $d_{1} \sim
d_{2} \ll R_{\rm{E}}$, where $R_{\rm{E}}$ is the Einstein
radius.)\footnote{The Einstein radius defines the radial scale for a
lensing configuration.  If a star lies exactly behind another one,
then due to the symmetry, a ring-like image appears.  This ring has an
angular radius $\theta_{\rm{E}}$ and a linear radius
$R_{\rm{E}} = d_{\rm{L}}\theta_{\rm{E}}$, where
$d_{\rm{L}}$ is the angular diameter distance from the observer
to the lens plane.  For distant galaxies acting as lenses, the
Einstein radius $R_{\rm{E}}$ is of order 1 arcsec.}  Hence the
authors concluded that when deriving an upper bound on
$R_{\rm{fold}}$, beyond which one can infer the presence of
small-scale structure, $d_{2}$ must also be taken into account along
with $d_{1}$.  The sensitivity of $R_{\rm{fold}}$ with respect to
the source's position along the fold caustic will unfortunately make
things difficult for us, and for this reason we will concentrate on
cusp points only, not fold points; see $\S 3$ below.

\begin{figure}
\begin{center}
\includegraphics[height=4cm,width=7cm]{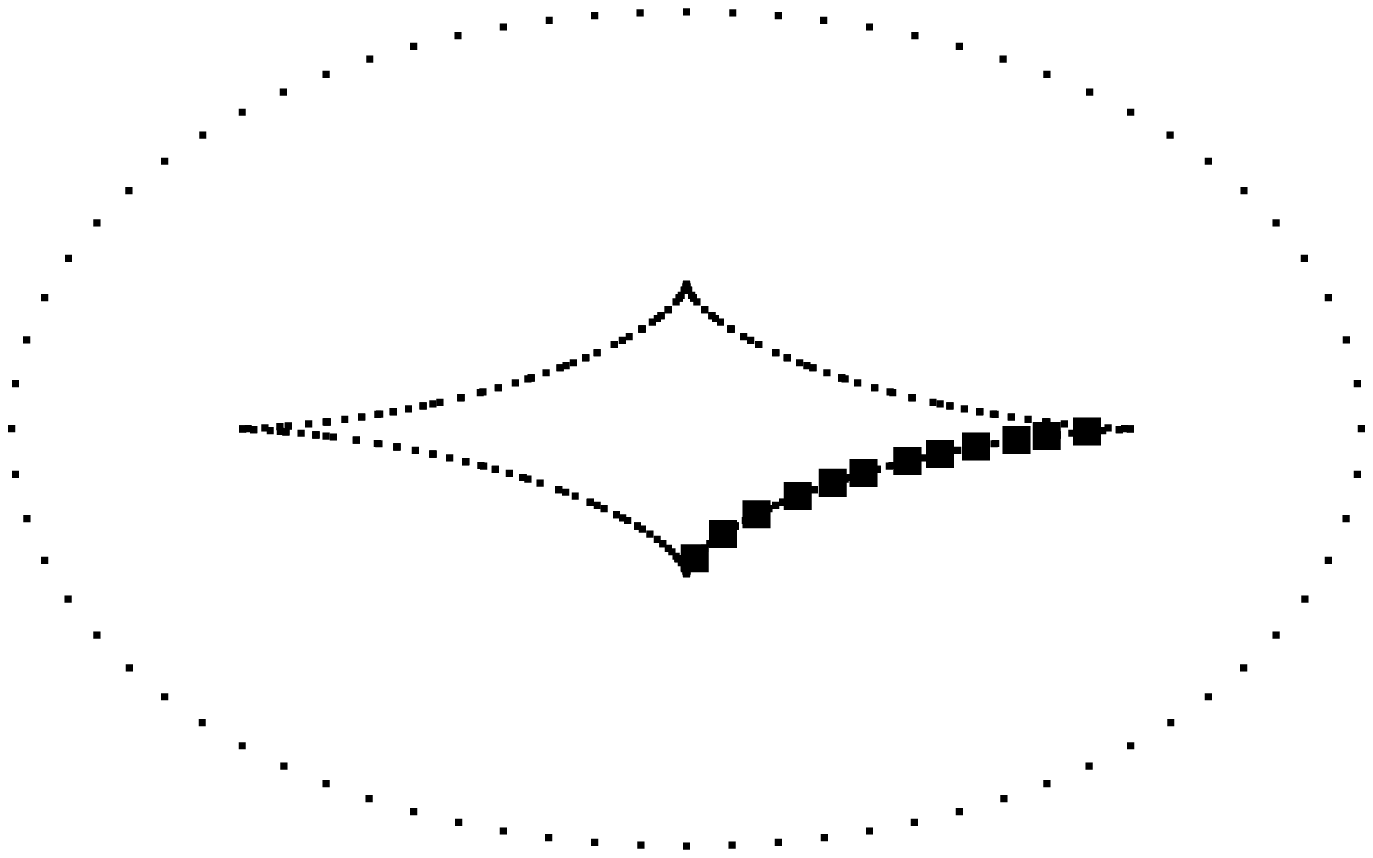}
\includegraphics[height=4cm,width=9cm]{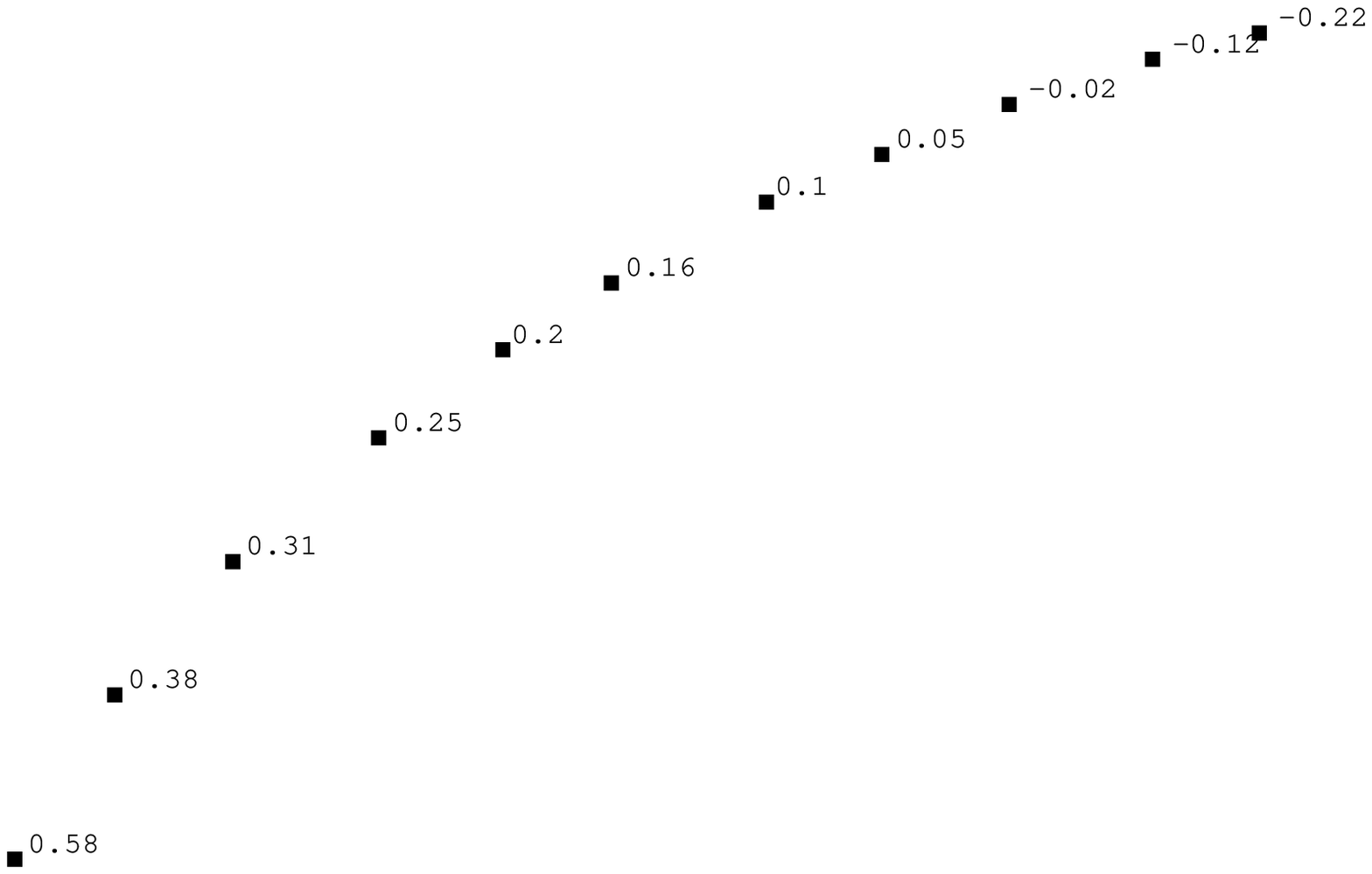}
\caption{Caustic curve for an SIE with ellipticity $e = 0.5$. The
numbers below are the values of $A_{\rm{fold}}$ in the asymptotic
fold relation $R_{\rm{fold}} = A_{\rm{fold}}d_{1}$,
evaluated at the points selected above. Note that $A_{\rm{fold}}$
is monotonically decreasing from left to right, as one moves from the
short axis to the long axis.  As Appendix A of Keeton et al. (2005)
discusses rigorously, $A_{\rm{fold}}$ is to be evaluated on the
caustic, but describes the fold relation in the vicinity of the
caustic.}
\end{center}
\end{figure}

Incidentally, in Figures 1 and 2, and in all of our calculations
below, we use as our lens model an SIE with ellipticity $e = 0.5$.
Although we are restricting our theoretical analysis to an SIE, our
results are nevertheless general, for the following reason.  Keeton et
al. (2005) generated a large ensemble of realistic lens potentials
drawn from three different observational samples, with varying values
of ellipticity, octupole modes, and external shear. For each lens
potential in their ensemble, they chose $\sim 10^{6}$ source
positions, solved the lens equation using Gravlens, and then computed
$(d_{1},d_{2},R_{\rm{fold}})$ for each minimum/saddle image pair.
They then plotted $R_{\rm{fold}}$ on the $(d_{1},d_{2})$ plane.
This enabled them to extract the probability distribution of
$R_{\rm{fold}}$ for fixed values of $d_{1}$ and $d_{2}$.  Next,
using an SIE with $e = 0.5$, they chose source positions such that
$d_{1}^{\rm{fold}} = d_{1}^{\rm{cusp}} =
.46R_{\rm{ein}}$, and calculated $R_{\rm{fold}}$.  They
compared this value of $R_{\rm{fold}}$ to that of each of the
lens potentials in their ensemble, for the same values of $d_{1}$ and
$d_{2}$.  Finally, for each minimum/saddle image pair, they found that
the value of $R_{\rm{fold}}$ calculated in the case of the SIE
fell {\it within} the probability distribution of their ensemble, and
thus concluded that each image pair of the SIE was consistent with
lensing by a realistic potential.  They repeated this procedure on the
23 known four-image lenses, and concluded that each case in which
$R_{\rm{fold}}$ lay {\it outside} the probability distribution
constituted strong evidence that small-scale structure was present for
that particular lens.  Given their result, we take as our {\lq
archetypal smooth lens potential\rq} an SIE with ellipticity $e =
0.5$.  However, we also vary the ellipticity and external shear and
evaluate their impact on our results.

\section{Investigating $R_{\rm{cusp}}$ and $R_{\rm{fold}}$}

Prior to adding our point-mass substructures to the SIE, we calculate
the values for $R_{\rm{cusp}}$ and $R_{\rm{fold}}$ for a
smooth lens model.  In what follows, we assume that the source sits a
very small but finite distance from the cusp or the fold.

For a source near a fold caustic but not near a cusp (mathematically,
this means that we pick a neighborhood around our source that does not
contain a cusp point), the image configuration is given in the top
panel of Fig.~1.  Calculating $R_{\rm{fold}}$ for the fold image
pair (A,B) that straddles the critical curve gives $R_{\rm{fold}}
\approx 0$ because, as stated in {\S 2}, A {\&} B have roughly equal
and opposite magnifications.  Of course we can apply
$R_{\rm{fold}}$ to {\it any} minimum/saddle image pair.  Now, the
two fold pair images A {\&} B will have much higher magnifications
than the other two images C {\&} D.  In fact, it can be as much as
three orders of magnitude greater.  So the fold relation for an image
in the fold image pair and one of the other two images will converge
to $R_{\rm{fold}} \approx \pm 1$.  If the image in the fold image
pair is a minimum, we get $+1$; if it is a saddle, we get $-1$.

The more interesting case is when a source lies near a cusp point, in
which case we can use both relations $R_{\rm{cusp}}$ and
$R_{\rm{fold}}$; see the middle panel of Fig.~1.  In this case
the magnifications of the outer two images A {\&} C will have the same
sign and exactly the same magnitude, while the magnification of the
middle image B will have the opposite sign and roughly twice the
magnitude, with the sign depending on whether the source lies near a
long axis cusp (B is a saddle) or near a short axis cusp (B is a
minimum).  Of course, as stated in $\S 2$, for the cusp triplet
(A,B,C), $R_{\rm{cusp}} \approx 0$.  We now apply
$R_{\rm{fold}}$ to the cusp triplet.  Pairing the middle image B
with either of the outer images A or C gives $R_{\rm{fold}}
\approx \pm 1/3$.  If B is a saddle, then A {\&} C are minima, so we
get $-1/3$; if B is a minimum, then A {\&} C are saddles, and we get
$+1/3$.  The fourth image D, not part of the cusp triplet, will have a
magnification much less than that of A,B, or C.  It can be as much as
three orders of magnitude less.  So the fold relation gives
$R_{\rm{fold}} \approx \pm 1$ for the combination of any one of
the cusp triplet images and D.  The sign will depend on which cusp
triplet image we use, and whether the source sits on the long or short
axis: for a long axis cusp, (A,D) and (C,D) give $+1$, while (B,D)
gives $-1$.

Of all these values, the important ones for our purposes are the
relations $R_{\rm{cusp}} \approx 0$ and $R_{\rm{fold}}
\approx \pm 1/3$ for the cusp triplet: as we will show in the next
section, substructure {\it breaks} the cusp triplet symmetry.  The
breaking of the symmetry causes the two outer images to no longer have
identical magnifications (and thus the same value for
$R_{\rm{fold}}$).  They are also no longer equidistant from the
middle image.
  
\section{Simulating the effects of substructure}

Since our examination involves a simulated configurations rather than
observational data, we can and choose to keep the source position
fixed and investigate the effect of substructure on the image
configurations.  Specifically, we investigate the change to both the
image configurations and to the values of $R_{\rm{fold}}$ and
$R_{\rm{cusp}}$ when we distribute substructure in the form of
point-masses onto our archetypal smooth lens potential, an SIE with
ellipticity $e = 0.5$.  We consider both random and symmetric spatial
distributions for 5-10 point-masses.  We examine cases when the
spatial distribution of these point-masses is (1) less than, (2)
roughly equal to, and (3) greater than the distance between the
images.  We use Gravlens to solve the modified lens equation (SIE +
point-masses) for a source placed very close to either a fold or cusp.
In each case, we gradually increase the mass of our point-masses from
$0$ (the control case) to the {\lq cut-off\rq} mass, which is the mass
value beyond which we no longer have a four-image lens.  Finally, we
calculate $R_{\rm{fold}}$ and $R_{\rm{cusp}}$ for these
granular lenses.

When considering sources near a cusp, we concentrate on long axis
cusps because they satisfy the ideal fold relation more readily than
short axis cusps (Keeton et al. 2005).  We also ignore cross
configurations (see the bottom panel of Fig.~1) because in such a case
$d_{1} \sim d_{2} \sim R_{\rm{E}}$, and scales on the order of
the Einstein radius are no longer {\lq small-scale\rq}.  (Recall from
$\S 2$ above that a violation of the cusp or fold relation implies the
presence of significant substructure on scales {\it smaller than the
distance between the images.}  Therefore calculating
$R_{\rm{fold}}$ for widely separated images does not really tell
us about small-scale structure, because we are no longer restricted to
short length scales.)

Of course, the shortest length scale (i.e. the smallest possible value
for $d_{1}$) is the distance between the fold image pair in a fold
image configuration (the two closely-spaced images in the top panel of
Fig.~1).  However, we encounter a problem when investigating sources
near a fold but not near a cusp.  In general, changing the lens
potential by adding substructure changes the shape of the caustic
curve, and the nature of the change depends on both the spatial
distribution of the substructure, and more importantly, their masses.
To give an example: setting the mass of our SIE to 1 (this sets
$R_{\rm{E}} = 1$ in the code units), the spatial distribution of
five point-masses in the manner shown in Fig.~3, with masses ranging
from $0 < m_{i} \leq 0.0190$, deforms the caustic curve considerably,
but still produces four-image lenses.  For masses $0.0190 \leq m_{i} <
0.0259$, the lens can also produce five images, though the fifth image
is quite faint: its magnification is $|\mu| \sim 10^{-5}$, while
$|\mu|$ for any one of the cusp triplet images is $\sim 10^{2}$.
Finally, for masses $m_{i} \geq 0.0259$, the lens can produce six
images.  Strictly speaking, the cusp and the fold relations are
defined only for four-image lenses, so we will not examine
$R_{\rm{cusp}}$ and $R_{\rm{fold}}$ in the five- and
six-image regimes.

We would like to come as close as we can to satisfying the ideal fold
and cusp relations, we want to place our source as close to the
caustic curve as Gravlens allows.  Now, aside from deforming certain
regions of the caustic curve, the addition of substructure also tends
to {\it narrow} the diamond-shaped region of the curve, so that a
source placed very close to a fold or cusp may actually end up falling
{\it outside} the diamond-shaped region of the {\it new} caustic
curve.  A source outside this region, of course, will not produce fold
or cusp configurations, so this means that we are forced to {\it move}
our source back inside the diamond-shaped region of our new caustic
curve in order to produce four-image configurations. We do encounter
the problem of trying to place our new source in the {\lq same
place\rq} with respect to the new fold caustic as our old source was
with respect to the old fold caustic.  Given the sensitivity of
$R_{\rm{fold}}$ to where our source lies along the fold (see \S 2
and Fig.~2), this is difficult in practice.  When we move our source
back inside the caustic curve, the value of $R_{\rm{fold}}$ will
change, but we cannot say with confidence whether this change in
$R_{\rm{fold}}$ is due to the addition of substructure, or simply
because we displaced our source by a small amount.  For this reason,
we ignore sources lying near a fold.

\begin{figure}
\begin{center}
\includegraphics[height=4cm,width=4cm]{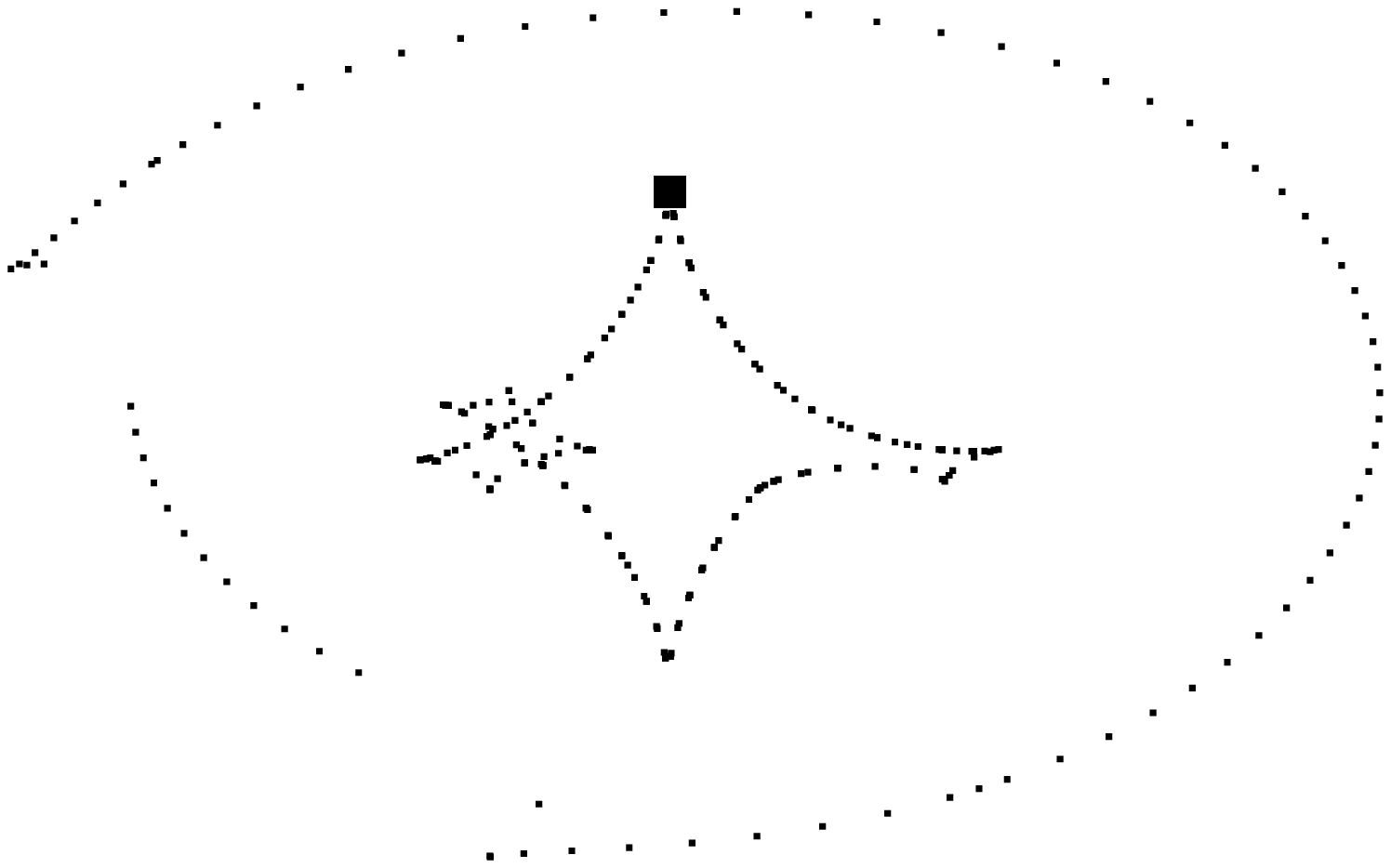}
\includegraphics[height=4cm,width=4cm]{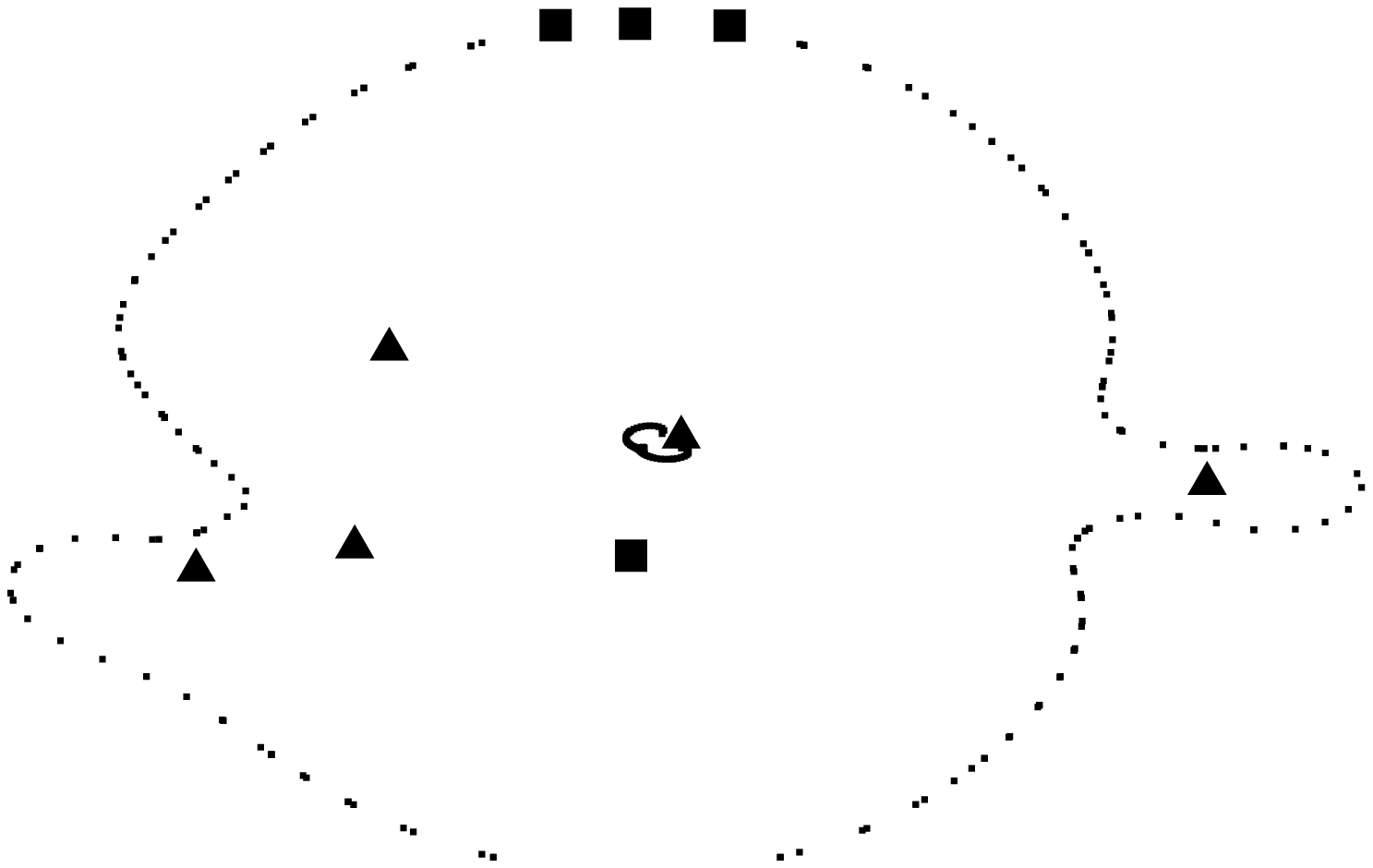}
\includegraphics[height=4cm,width=4cm]{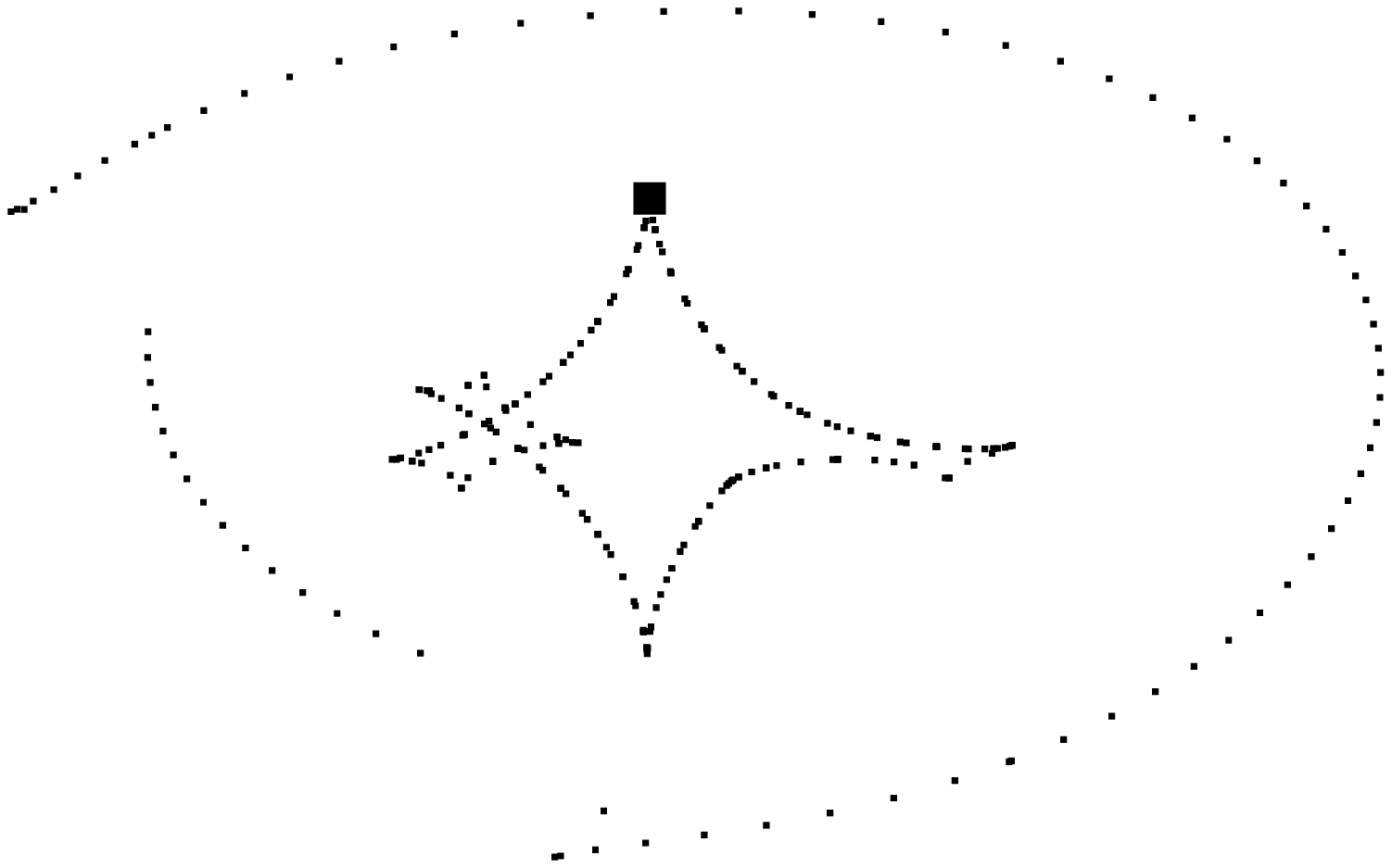}
\includegraphics[height=4cm,width=4cm]{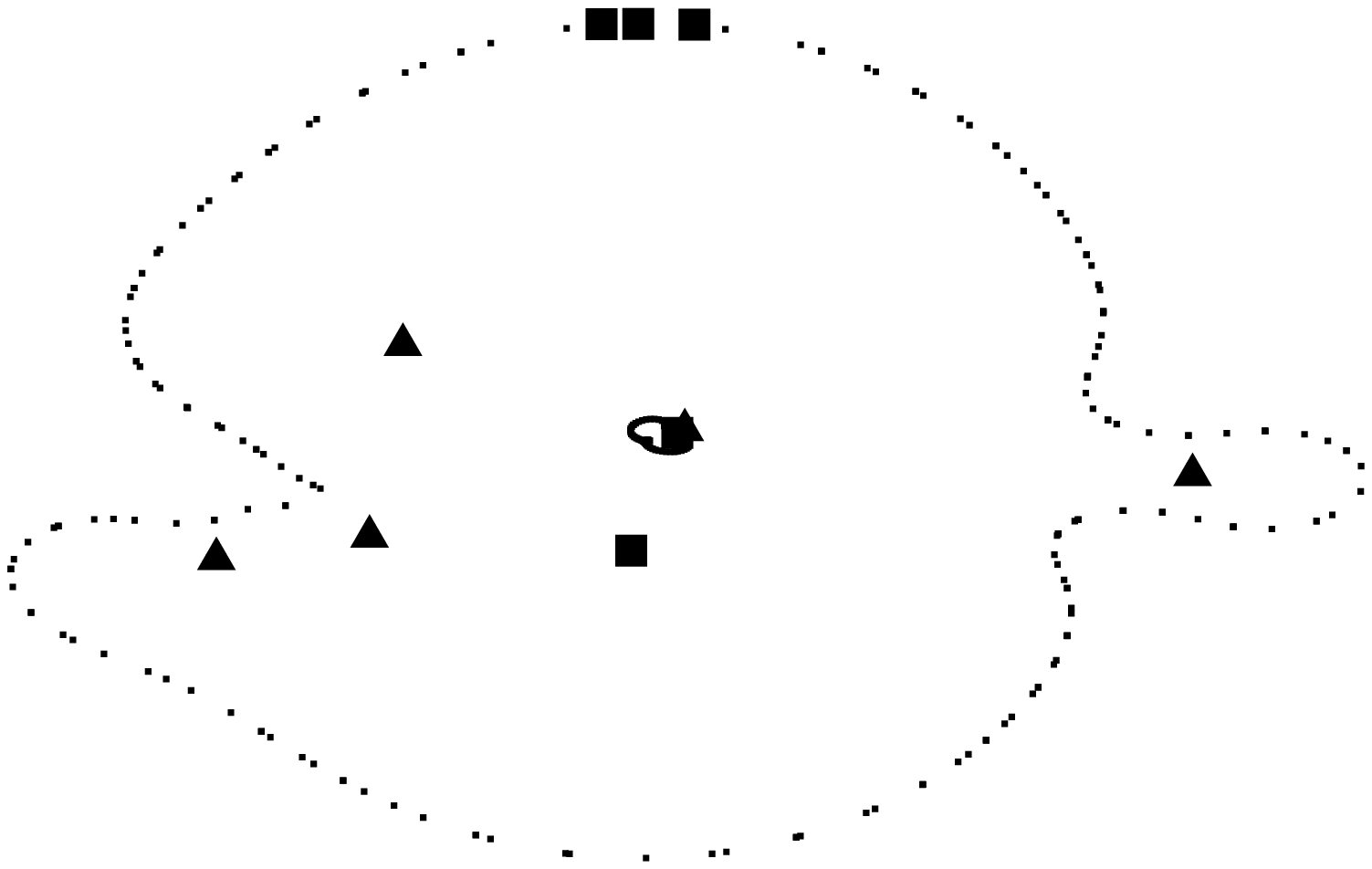}
\includegraphics[height=4cm,width=4cm]{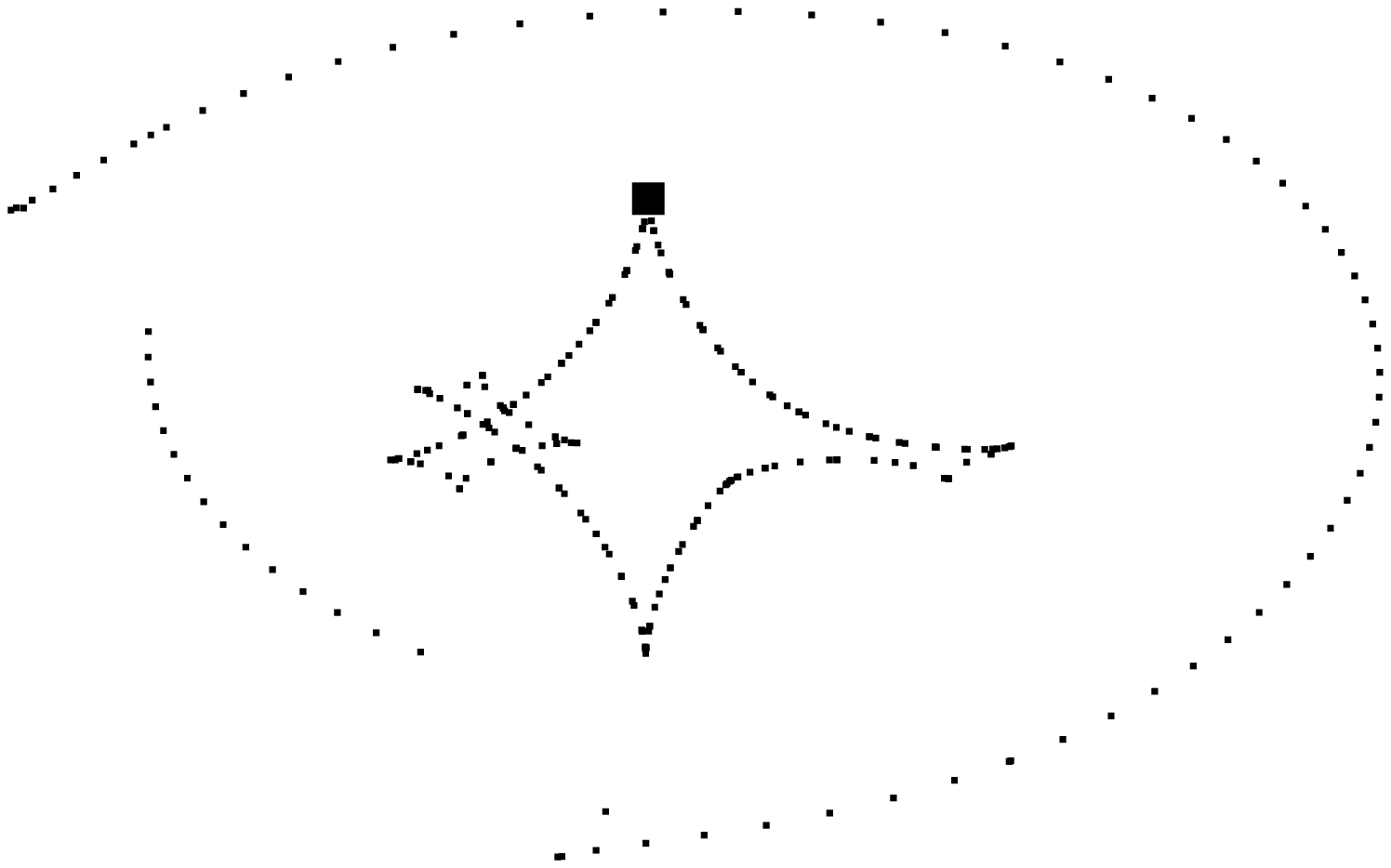}
\includegraphics[height=4cm,width=4cm]{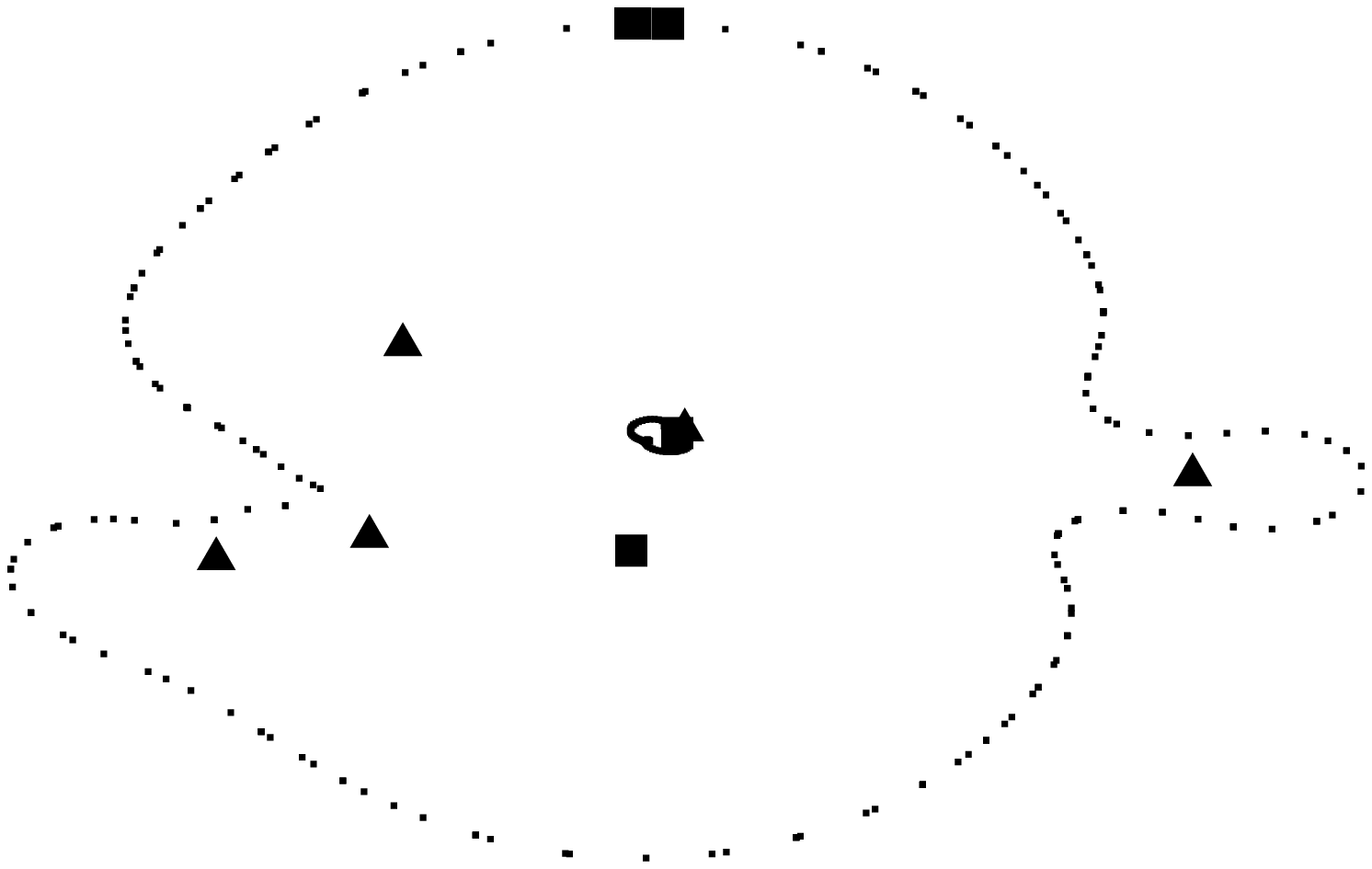}
\caption{The breaking of the cusp triplet symmetry for a distribution
of five point-masses onto an SIE with ellipticity $e = 0.5$.  In each
panel, the diagram on the left shows the caustic curve and source
position while that on the right shows the corresponding critical
curve and cusp configuration.  For each caustic curve we have moved
our source as close to the cusp as Gravlens allows.  In each case,
note the deformations of both the caustic and critical curves due to
the point-masses, and compare these with the cusp configuration in
Fig.~1.  The triangles are the point-masses and the squares are the
images. In each panel, the point-masses have the same mass.  In the
first panel, that mass is $m = 0.0185$ (in units where
$R_{\rm{E}} = 1$); in the second panel, $m = 0.0255$; in the
third, $m = 0.0258$.  Note the fifth image near the origin in the
second and third rows.  For $m \geq 0.0259$, we get six images.}
\end{center}
\end{figure}

Fortunately, we do not encounter this problem when we place our source
near a {\it cusp}, because the cusp is a much smaller region than the
fold caustic.  We cannot {\lq move along\rq} the cusp in the same way
that we can move along the fold because the cusp is essentially a tiny
wedge.  Therefore we can safely move our source back inside the
caustic curve and place it back into the wedge of the cusp.  We still
have to be careful, however, because the addition of substructure
tends to displace the long axis cusp slightly off the axis (so that
the long axis is no longer aligned with the coordinate axis), making
it more difficult to locate the cusp accurately.  Doing so gives us
our most interesting result: when we add substructure onto our SIE,
and consider a source near our new long axis cusp, we no longer have
the {\it symmetry} in our cusp triplet.  The cusp triplet tends to be
displaced to one side, so that the two outer images are no longer
identical.  In fact, one of them is closer to the middle image.  This
is illustrated in the three panels shown in Fig.~3.  At this point an
important issue arises: because the cusp point is more difficult to
locate when substructure displaces the long axis, one may wonder
whether the skewed cusp triplet shown in Fig.~3 is caused merely by
the possibility that we have missed the cusp point and instead placed
our source against a {\it fold} caustic.  If so, then we would be able
to reproduce the same skewed cusp triplet in the case of an SIE {\it
without} substructure, simply by displacing our source off the cusp by
a small amount.  But in fact this is not the case: a source displaced
slightly off the cusp for an SIE without substructure produces a very
tight fold configuration (as in the top panel of Fig.~1), one that
looks altogether different from the skewed cusp triplet shown in
Fig.~3.  Thus it is the {\it substructure} that breaks the cusp
triplet symmetry.

\section{Results and Conclusions}

We now examine the sensitivity of the cusp and the fold relations to
the spatial distribution of substructure shown in Fig.~3. We set the
mass of our SIE to 1 ($R_{\rm{E}} = 1$) and vary the substructure
masses from $0$ (the control case) to $m_{i} = 0.019$. For a typical
galaxy lens this corresponds to adding substructure in the mass range
of 10$^{5}$ - 10$^{7}\,M_{\odot}$. Note that for $m_{i} > 0.019$ we no
longer have a four-image lens.

\subsection{Dependence on Substructure Mass}

First of all, for the control case of the smooth potential and no
substructure, we find $R_{\rm{cusp}} = 0.002$ and
$R_{\rm{fold}} = (-0.331,-0.331)$, where $R_{\rm{fold}} =
-0.331$ for both the left/middle image pair and the right/middle image
pair.  As discussed in $\S 3$, we expect a value close to $0$ for
$R_{\rm{cusp}}$ because we have placed our source as close to the
cusp as Gravlens allows (see Fig.~1).  However, our source still sits
a finite distance from the cusp, not asymptotically close, so we
should not expect to satisfy the ideal cusp relation
$R_{\rm{cusp}} \to 0$ exactly; hence $R_{\rm{cusp}} = 0.002$
is acceptably close to zero.  As for $R_{\rm{fold}} = -0.331$,
this value, too, is what we expect: both the leftmost and rightmost
images have the same magnification, which is half the magnitude as
that of the middle image with the opposite sign.  In this case our
source sits near a long axis cusp, so the middle image is a saddle and
$R_{\rm{fold}}$ is negative.  As discussed in $\S 3$, we expect a
value of $R_{\rm{fold}} \approx -0.333$, so a value of $-0.331$
is close enough.

Now we examine the breaking of the cusp triplet symmetry shown in
Fig.~3, for three mass values in the four-image regime:
\begin{itemize}
  \item $m = .000$ : $R_{\rm{cusp}} = .002$ and $R_{\rm{fold}} = (-.331, -.331)$
  \item $m = .001$ : $R_{\rm{cusp}} = .005$ and $R_{\rm{fold}} = (-.326, -.332)$
  \item $m = .005$ : $R_{\rm{cusp}} = .006$ and $R_{\rm{fold}} = (-.317, -.340)$
  \item $m = .010$ : $R_{\rm{cusp}} = .006$ and $R_{\rm{fold}} = (-.305, -.352)$
  \end{itemize}
The notation $R_{\rm{fold}} = (-0.317, -0.340)$ implies that the
left/middle image pair gives $R_{\rm{fold}} = -0.317$, while the
right/middle image pair gives $R_{\rm{fold}} = -0.340$; likewise
for the others.  The change in $R_{\rm{fold}}$ manifestly reveals
the breaking of the cusp triplet symmetry; the change in
$R_{\rm{cusp}}$, however, does {\it not,} as shown in the top
panel of Fig.~4.  In Fig.~4 we show the effect of varying the mass of
the substructure on $R_{\rm{cusp}}$ and $R_{\rm{fold}}$.  It
is clear that, within the four-image regime, $R_{\rm{fold}}$
readily reflects the change in the mass of the substructure, while
$R_{\rm{cusp}}$ stays roughly constant.

\begin{figure}
\begin{center}
\includegraphics[width=8cm]{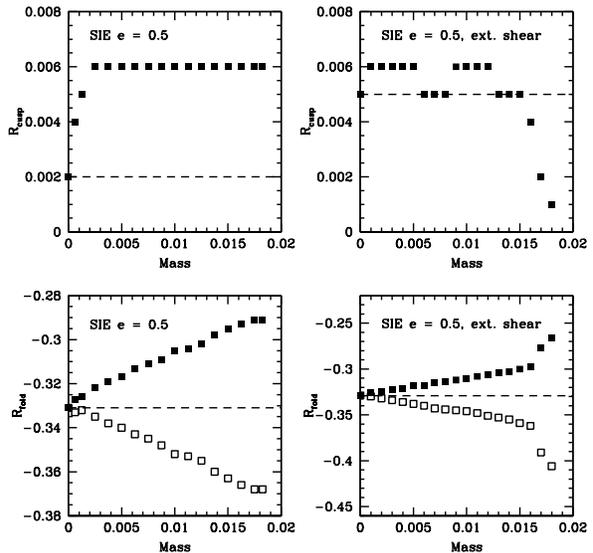}
\caption{$R_{\rm{cusp}}$ and $R_{\rm{fold}}$ as a function
of the mass $m$ of the substructure for the $e = 0.5$ case, for the
spatial distribution given in Fig.~3.  {\it Left column:} Without
shear.  For $0 \leq m \leq 0.0185$ (in units where $R_{\rm{E}} =
1$), we get four images for a source as near the long axis cusp as
Gravlens allows.  {\it Right column:} External shear with an amplitude
of $\gamma \sim 5\%$.  In this case we have a four-image regime for $0
\leq m \leq 0.020$.  The control case (no substructure) is shown by
the dashed line.  In the bottom panel, the black and white data points
denote the left/middle and right/middle image pairs in the cusp
triplet, respectively. $R_{\rm{fold}}$ reflects the change in the
mass of the substructure better than $R_{\rm{cusp}}$.}
\end{center}
\end{figure}

Keeton et al. (2005) demonstrated that, for observed cusp lenses, the
fold relation indicated the same flux ratio anomalies as the cusp
relation, with one exception: in B2045+265, the possible presence of
octupole modes precluded the authors from declaring that the fold
relation was violated, whereas the cusp relation was very clearly
violated.  However, because the fold relation is defined for pairs, it
gave the additional information of indicating {\it which particular
image} in the cusp triplet was the one most affected by substructure,
a distinction that could not be made with the cusp relation.  What we
have shown in our analytic argument here is that, for an SIE with $e =
0.5$, the fold relation is a more reliable indicator of substructure
than the cusp relation.

\subsection{Sensitivity to External Shear}

Both $R_{\rm{cusp}}$ and $R_{\rm{fold}}$ are expected to be
sensitive to the presence of external shear.  We therefore add
external shear with an amplitude of $\gamma \sim 5\%$ to the
configurations.  With this modification, we examine the breaking of
the cusp triplet symmetry for the same substructure mass values listed
in {\S 4.1}, and find:
\begin{itemize}
  \item $m = .000$ : $R_{\rm{cusp}} = .005$ and $R_{\rm{fold}} = (-.329, -.329)$
  \item $m = .001$ : $R_{\rm{cusp}} = .006$ and $R_{\rm{fold}} = (-.326, -.330)$
  \item $m = .005$ : $R_{\rm{cusp}} = .006$ and $R_{\rm{fold}} = (-.318, -.338)$
  \item $m = .010$ : $R_{\rm{cusp}} = .006$ and $R_{\rm{fold}} = (-.311, -.346)$
\end{itemize}
Once again, $R_{\rm{fold}}$ manifestly reveals the breaking of
the cusp triplet symmetry, whereas $R_{\rm{cusp}}$ does not.  In
fact, as Fig.~4 shows, $R_{\rm{cusp}}$ eventually falls {\it
below} the control case value of $R_{\rm{cusp}} = 0.005$.  On the
other hand, $R_{\rm{fold}}$ remains correlated with the mass of
the substructure throughout the four-image regime, even with external
shear. Finally, note that the four-image regime in this case is
produced when $m_{i} < 0.02$.

\subsection{Sensitivity to Ellipticity}

The ellipticity of our SIE affects the values of $R_{\rm{cusp}}$
and $R_{\rm{fold}}$.  We find that for ellipticities lower than
our archetypal value of $e = 0.5$, $R_{\rm{cusp}}$ remains
unresponsive to the substructure, whereas $R_{\rm{fold}}$ is
still sensitive to the mass of substructure, though not as robustly as
for the case with $e = 0.5$.  The left-hand panels of Fig.~5
demonstrates this explicitly for an ellipticity $e = 0.25$.  For
ellipticities higher than our archetypal value, $R_{\rm{cusp}}$
displays erratic fluctuations, whereas $R_{\rm{fold}}$ remains
well correlated with the mass of the substructure, as shown in the
right-hand panels of Fig.~5.

\begin{figure}
\begin{center}
\includegraphics[width=8cm]{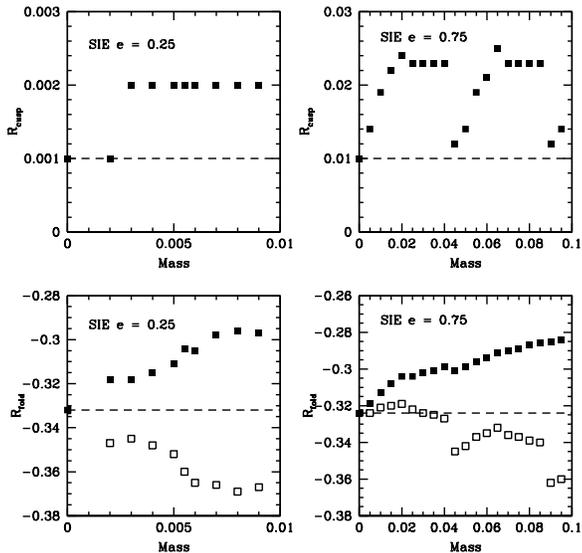}
\caption{{\it Left column:} $R_{\rm cusp}$ and $R_{\rm fold}$ as in
Fig.~ 4, but now for a lens with ellipticity $e = 0.25$. We have a
four-image regime for $0 \leq m \leq 0.009$.  We ignore the region $0
< m < 0.002$ because it is difficult to cleanly place the source near
the cusp point.  {\it Right column:} The relations for a lens
ellipticity $e = 0.75$.  In this case we have a three-image regime (a
cusp triplet, no fourth image) for $0 \leq m \leq 0.035$, and
four-image regime for $0.035 < m \leq 0.100$.  Note that although
$R_{\rm{cusp}}$ fluctuates erratically, in this case the
right/middle image pair of $R_{\rm{fold}}$ is not reliable,
either.  The most reliable indicator of the substructure masses here
is the left/middle image pair of $R_{\rm{fold}}$.}
\end{center}
\end{figure}

\subsection{Concluding remarks}

In this theoretical investigation, we attempt to develop a robust
diagnostic that quantifies the presence of substructure in lens
galaxies.  To this end, we simulated a simple lens potential with
added substructure, with the aim of seeing how the cusp and the fold
relations, calculated while keeping the source position fixed, respond
to the presence of the substructure as we vary its mass and position
within the lens.  We took as our lens model an SIE and used
point-masses as our substructure with masses ranging from $0 < m_{i}
\leq 0.019$ ($R_{\rm{E}} = 1$).  When we varied the ellipticity
of our SIE, we found that for low ellipticities the cusp relation was
unresponsive to the substructure, whereas for high ellipticities it
became erratic.  The fold relation, on the other hand, remained well
correlated with the mass of the substructure.  Considering the effect
of external shear gave the same result: the fold relation remained
well correlated with the substructure, whereas the cusp relation did
not.

We considered random distributions of point-masses as substructure and
cases when their spatial distribution was less than, roughly equal to,
and greater than the distance between the images.  Overall, we found
that the addition of substructure breaks the symmetry of the cusp
triplet, and that the fold relation responds more accurately to the
change in mass of the substructure than the cusp relation.  We
conclude, therefore, that the fold relation is the more robust
diagnostic of substructure.

In order to apply this technique to real data, we will need to use the
observed image positions and the distances between them, and then use
Monte Carlo methods to constrain the corresponding source positions.
While this is beyond the scope of this work, we pursue it in a
follow-up paper.

\section*{Acknowledgments} 

We thank Scott Gaudi for a careful reading of the manuscript and
constructive comments; Charles Keeton for his publicly available
Gravlens software and Arlie Petters for helpful advice.  ABA also
thanks George Mias for help with the figures.

\end{document}